# The Rapid Imaging Planetary Spectrograph

Patrick Lierle[1,2], Carl Schmidt[2], Jeffrey Baumgardner[2], Luke Moore[2], Emma Lovett[2]

[2]*Boston University, Boston, MA, USA*

Abstract:

The Rapid Imaging Planetary Spectrograph (RIPS) was designed as a long-slit high-resolution spectrograph for the specific application of studying atmospheres of spatially extended solar system bodies. With heritage in terrestrial airglow instruments, RIPS uses an echelle grating and order-sorting filters to obtain optical spectra at resolving powers up to R~127,000. An ultra-narrowband image from the reflective slit jaws is captured concurrently with each spectrum on the same EMCCD detector. The "rapid" portion of RIPS' moniker stems from its ability to capture high frame rate data streams, which enables the established technique known as "lucky imaging" to be extended to spatially resolved spectroscopy. Resonantly scattered emission lines of alkali metals, in particular, are sufficiently bright to be measured in short integration times. RIPS has mapped the distributions of Na and K emissions in Mercury's tenuous exosphere, which exhibit dynamic behavior coupled to the planet's plasma and meteoroid environment. An important application is daylight observations of Mercury at solar telescopes since synoptic context on the exosphere's distribution comprises valuable ground-based support for the upcoming BepiColombo orbital mission. As a conventional long slit spectrograph, RIPS has targeted the Moon's surface-bound exosphere where structure in linewidth and brightness as a function of tangent altitude are observed. At the Galilean moons, RIPS can study the plasma interaction with Io and place new constraints on the sputtered atmosphere of Europa, which in turn provides insight into the salinity of Europa's subsurface ocean. The instrumental design and construction are described herein, and these astronomical observations are presented to illustrate RIPS' performance as a visiting instrument at three different telescope facilities.

---

[1] Corresponding author plierle@bu.edu
[2] Center for Space Physics, Boston University, 725 Commonwealth Ave, Boston MA 02215, USA

# I. Design Motivation

## I.1 Existing Instrumentation

The Rapid Imaging Planetary Spectrograph (RIPS) is a long-slit high-resolution spectrograph designed for studying the atmospheres of spatially extended solar system bodies. RIPS uses an echelle grating and order-sorting filters to obtain optical spectra at resolving powers up to R~127,000. An ultra-narrowband image from the reflective slit jaws is captured concurrently with each spectrum on the same EMCCD detector. This allows RIPS to apply the technique known as "lucky imaging" to spatially resolved spectroscopy. By combining spectra only where atmospheric turbulence is momentarily still—as determined from the synchronized narrowband images—RIPS is capable of generating high-resolution spatial maps of gases in planetary atmospheres.

RIPS draws inspiration from the prolific line of echelle spectrographs that came before it. Perhaps the best known long-slit high-resolution spectrograph is the HIgh-Resolution Echelle Spectrograph (HIRES) at Keck I on Mauna Kea, Hawaii. HIRES is a grating cross-dispersed echelle spectrograph operating between 3,100 and 10,000Å (Vogt et al., 1994). Its primary application is stellar radial velocity measurements to characterize exoplanet orbits. Slit widths between 0.4″ and 1.7″ correspond to resolving powers of 25,000 to 87,000. Fixed HIRES slit lengths up to 28″ are generally chosen to optimize between spatial information and desired wavelength. With a short 3.5″ slit length, blue echelle orders are fully separable on its 3 detectors, while longer slits lengths have some order overlap. The Ultraviolet and Visual Echelle Spectrograph (UVES) is an analogous cross-dispersed echelle spectrograph with variable slit length mounted at the Very Large Telescope in Chile (Dekker et al., 2000). McDonald Observatory in Fort Davis, Texas, is home to the cross-dispersed Tull Spectrograph, which operates in four different configurations at the Coudé focus of the 2.7m Harlan J. Smith Telescope to attain resolutions up to R=240,000 (Tull et al., 1995).

Instrumentation in this field has increasingly moved to fiber-fed stabilized spectrographs for the point source targets, and several well-known long-slit echelle spectrographs now have been decommissioned. SARG at the Italian Telescopio Nazionale Galileo offered extremely high resolution of up to R=164,000 for wavelength range of 3,700 to 10,000Å (Gratton et al, 2001). The R~65,000 Cassegrain Echelle Spectrograph (CASPEC) was an important early instrument at the European Southern Observatory's 3.6m Telescope in La Silla, Chile (D'Odorico et al., 1983). SARG and CASPEC were superseded by the fiber-fed HARPS North and South instruments dedicated to precision radial velocity measurements for the exoplanet community. Also decommissioned is the 'LPL Echelle,' a Cassegrain "quasi-Littrow" echelle spectrograph with a resolving power of R=160,000 used to spatially resolve emissions in planetary atmospheres and their moons from the Steward Observatory 61" Kuiper Telescope in Arizona (Hunten et al., 1991). Precision radial velocity has been a major driver in new instrumentation and the decommissioning of these long-slit instruments has left fewer tools available.

## I.2 Applications within Planetary Science

Long-slit high-resolution spectroscopy resolves emission features both spatially and spectrally. It is a well-established technique for investigating solar system atmospheres because these objects are generally bright and spatially extended, and their observing geometries can be optimized to distinguish between telluric and extraterrestrial features using planetary Doppler shifts. An overview of these applications helps set the context for RIPS' potential science applications, but the examples herein are by no means an exhaustive list.

Many of the earliest examples of the technique are studies of the rich optical emission spectra of cometary comae. The Tull Spectrograph on the 2.7-m telescope of McDonald Observatory once identified more than 12,000 individual emission lines in a single cometary coma, and thousands more remain unidentified (Cochran & Cochran, 2002). High spectral resolution is critical not just to separate telluric and extraterrestrial emission lines, but also to separate lines that would otherwise be blended, and to distinguish gas emission from the dust continuum. Cometary ions are picked up by the solar wind, making Doppler shift valuable to isolate neutral and ionic emissions. Long-slit high-resolution spectroscopy is an important tool to understand compositional differences in comet populations. The Tull Spectrograph has also offered the longest running survey of cometary comae; notably, Cochran et al. (2012) characterized the composition of 130 comets between 1980 to 2008. UVES and HIRES instruments operate from high altitudes where Earth's atmosphere can pass OH and OD emissions in the near-ultraviolet (Hutsemékers et al., 2008a). The spatial distributions in a comet's coma give information about the radiation pressure acting on the dust particles, and in the gas emission. A feedback relationship between solar Doppler shift, emission and radiation pressure was first identified by Swings (1941) and Greenstein (1958).

Long-slit high-resolution spectroscopy also has a rich history of discoveries at the Galilean satellites. Io's atmosphere is lost at rates of more than a ton per second, and the resultant oxygen 6300Å cloud encircles Jupiter along the satellite's orbital path (Brown, 1981; Thomas 1996). The broad linewidth of sodium D emissions offered the first evidence of a strong coupling between Io and the Jovian magnetosphere (Brown & Chaffee, 1974; McElroy, Yung, and Brown, 1974), and the lines are duplicated at slow and fast Doppler shifts illuminating different atmospheric escape pathways by Jean's escape and ion recombination, respectively (e.g., Cremonese et al. 1992). Thomas (1996) reported a similar fast component of Io's K 7699Å cloud in observations from CASPEC. SARG measurements have also identified other Na Doppler shifts remote from Io, interpreted as negatively charged dust grains lofted by volcanic plumes (Grava et al. 2021).

Brown & Hill (1996) used HIRES to discover an extended sodium exosphere at Europa, followed by the Brown (2001) measurement of potassium. While both species rapidly escape Io, the Na/K ratio at neighboring Europa was discordant with Iogenic sources, suggesting these alkalis might instead originate from a salty subsurface ocean (F Leblanc et al., 2002). NaCl absorption features were then identified by the Hubble Space Telescope to show that this salt is localized to Europa's chaos terrain (Trumbo et al., 2019a). High salinity brines better retain a liquid phase, consistent with an especially shallow ice layer and/or geologically recent melting as explanations for Europa's chaos terrain topology (B E Schmidt et al., 2011; Steinbrügge et al., 2020a).

Near quadrature, the Galilean satellites can be observed during passage through Jupiter's shadow, which enables a number of interesting experiments. In the absence of sunlight, emissions are auroral in nature, and the Io plasma torus offers an abundant supply of electrons to excite gas transitions in the satellite atmospheres. High resolution is needed to separate such emissions from lines in Earth's atmosphere, and long-slit spectroscopy is critical for successfully removing the bright background spectrum to isolate faint emission only an arcminute from Jupiter's sunlit disk. Grava et al. (2014) observed a factor of ~4 decrease in total atomic sodium content when Io emerged from Jupiter's shadow with SARG, which Schmidt et al. (2023) showed is due to the solar control of the $NaCl^+$ ion. HIRES was used to discover oxygen 6300Å aurora at Europa in Jupiter's shadow (de Kleer and Brown, 2018a) and this technique recently revealed new colors of optical aurora at all four Galilean satellites (de Kleer et al. 2023; Schmidt et al. 2023).

Like Europa, the atmospheres of the Moon and Mercury are collisionless surface-bound exospheres. Sodium and potassium gases were discovered at these bodies using echelle spectrographs in the mid-1980s (Potter and Morgan, 1985; 1986; 1988, Tyler et al. 1988). The Moon is an easy astronomical target, but its gas emission is faint in contrast to intense moonlight, so the exosphere is only observable above the limb. Sampling is also limited to the same limb coordinates because our Moon is tidally locked. This makes it difficult to disentangle orbital effects like passage through the Earth's magnetotail from the local effects produced by the heterogeneous chemical composition of the lunar surface. While Mercury's surface brightness is the highest among all bodies mentioned herein, its exospheric emission is amply bright to be isolated above the bright disk. Mercury's proximity to the Sun, however, limits viewing to observations in daylight or during narrow time windows at high airmass in twilight.

**I.3 Observational Considerations Unique to Mercury**

Mercury's unique set of challenges heavily motivated the RIPS design. At twilight, high airmass observations inherently suffer from non-photometric sky conditions, differential atmospheric refraction, suboptimal seeing, poor telescope tracking, and wind shake. Forward modelling where flux calibration is achieved using the planet itself as a photometric reference can address non-photometric skies. Differential refraction is addressed by filtering the slit jaw image to the same wavelengths that the spectrum samples. High-cadence imaging can in part mitigate the other undesirable effects. Short integration times can achieve imaging and high-resolution spectroscopy with reasonable signal to noise because Mercury is an exceptionally bright target, often reaching negative magnitudes. Blurring due to wind shake and tracking instability is reduced in shorter integrations. "Lucky imaging" is an established technique to mitigate atmospheric seeing by discarding the majority of frames and keeping only those where Earth's atmospheric turbulence is momentarily still (David L. Fried, 1978). All planetary science targets we have discussed so far are well served by conventional long-slit spectroscopy, but Mercury is sufficiently bright that the principles of lucky imaging can be extended to high resolution spectroscopy with significant advantages.

Even during optimal solar elongations, twilight observing windows may have insufficient duration to capture dynamics in Mercury's exosphere, and monitoring the planet for longer than ~40 min intervals requires daylight observations. This observation requires extreme care given Mercury's proximity to the Sun and the hazards of off-axis focused sunlight. With the closure of McMath-Pierce, the only major solar telescopes presently confirmed to allow daylight observations of Mercury are THEMIS at Tenerife and the Dunn Solar Telescope in Sunspot, NM. Leblanc et al. (2008; 2009) used THEMIS to measure variability in Mercury's polar emissions on hourly timescales, presumably due to changing solar wind conditions and plasma precipitation through the magnetic cusps. Additionally, THEMIS has detected dramatic changes during the passage of interplanetary coronal mass ejections, with the important implication that ground-based observations of the Na exosphere can serve as a remote proxy for the space weather conditions at Mercury (Orsini et al., 2018).

The extended observing periods allowed by solar telescopes enable RIPS to offer ground-based support of the upcoming BepiColombo mission (orbital insertion 5 December 2025). The two BepiColombo spacecraft will measure the solar wind and exosphere simultaneously. Ground-based exosphere comparisons would benefit two instruments in particular: the Mercury Sodium Atmospheric Spectral Imager (MSASI; Yoshikawa et al., 2010), a Fabry-Perot spectrometer dedicated to the Na D2 line, and the Probing of Hermean Exosphere By Ultraviolet Spectroscopy instrument (PHEBUS; Quémerais et al., 2020), a UV spectrometer with two dedicated channels for visible emissions.

## II. Design

### II.1 Optical Design

Motivated by these applications in planetary spectroscopy, the RIPS design combines a traditional high-resolution long-slit spectrograph with the concept of lucky imaging. To aid in image quality registration, spatial registration, and flux calibration, the narrowband slit image and the spectrum are imaged simultaneously on one detector. Figure 1 illustrates the RIPS optical layout. Telescope light enters the instrument from the left through a circular aperture capable of accommodating a 2" pre-filter, which can be used to reduce continuum light entering the instrument. The beam reflected by the slit jaws encounters steering mirror M1 and passes through a five-position, commercial-off-the-shelf (COTS) filter wheel housing anti-reflection coated Kodak Wratten 2 neutral density filters. This allows for brightness adjustment between the spectral and imaging channels; since both the slit image and the spectrum are captured on the same detector, a single integration time must be used for both. One filter position provides a neutral density filter 'sandwiched' with a narrowband (4 to 13Å) filter to eliminate the effects of differential atmospheric refraction resulting in a spatial shift between the slit image and the spectrum (if no filter is used in front of the slit). After the filter wheel, steering mirrors M2 and M3 direct the beam through a motorized pair of achromatic re-imaging lenses and position a focused image of the slit jaws on the bottom half of the Andor iXon Ultra 888 1024 x 1024 EMCCD (Model No.: DU-888U3-CS0-#BV).

Light passing through the slit aperture has a separate optical pathway. This beam continues to a flat folding mirror that re-directs the diverging light into a 900mm focal length, F/9 achromatic collimating lens. RIPS has two reflection grating options that are easily interchangeable: a 31.6 lines/mm, R/2, 110mm x 220mm, 62.5° blaze angle echelle, and a low dispersion 600 lines/mm first-order grating. RIPS operates as a Littrow spectrograph (with the angle between the incident and the diffracted beam being only 1.5°), using the same collimating lens as an objective lens to focus light dispersed by the grating. This light path then travels back to the folding mirror before it reaches M4, the first of four steering mirrors used to direct the image of the spectrum onto the EMCCD. An aerial image is formed between M4 and M5. The spectrum is re-imaged onto the top half of the detector after passing through a filter wheel, steering mirror M6, two achromats and a final steering mirror, M7—just like on the imaging side. In this arrangement, the grating is oriented along the blaze angle for maximum efficiency.

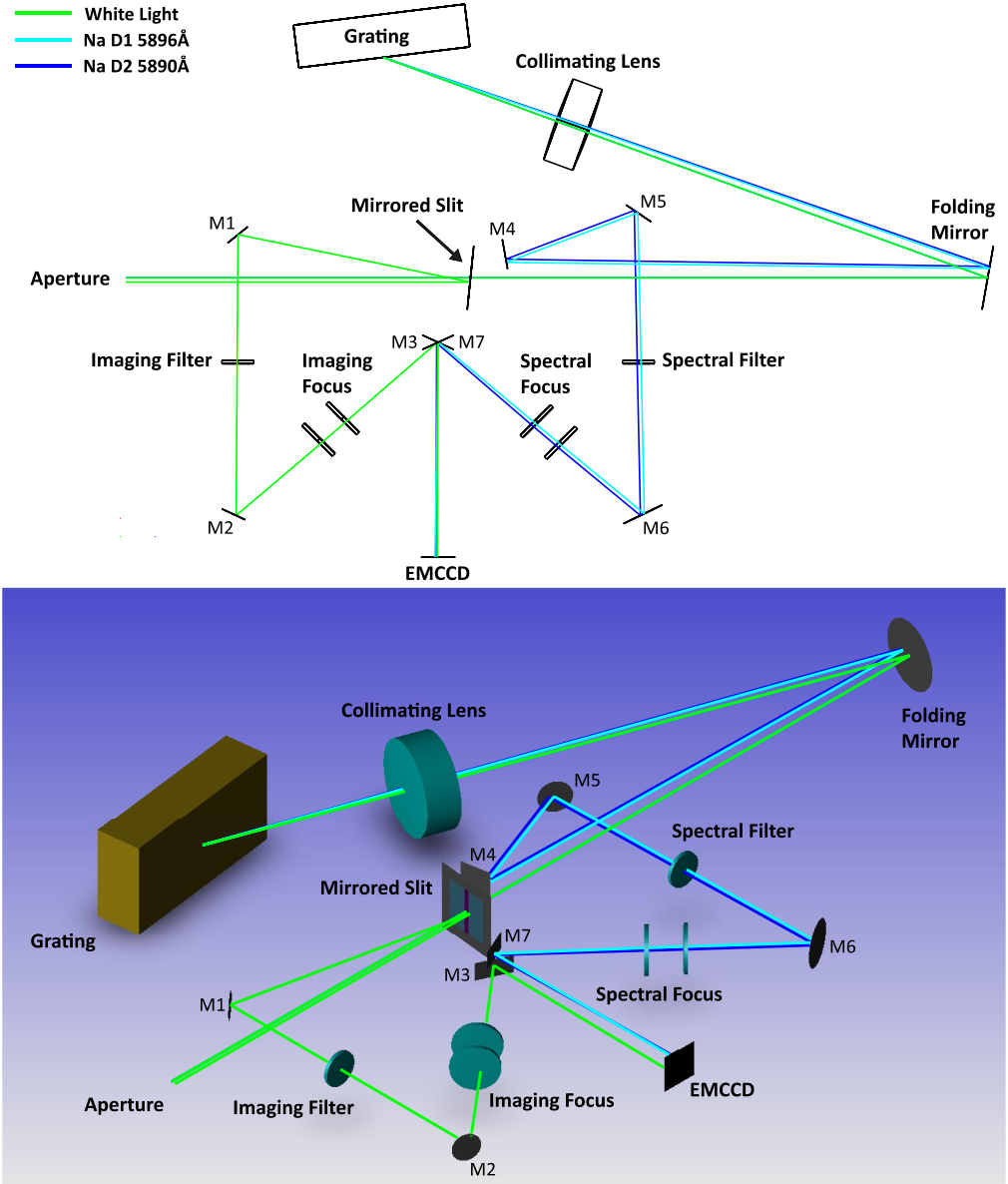

**Figure 1:** *Top:* A top-down optical design layout of RIPS. *Bottom:* A 3D Zemax OpticStudio rendering of the RIPS optical layout.

The spectral channel filter wheel contains four narrowband order-sorting filters for commonly studied emission lines in planetary atmospheres—Na, K, O, and S II—as well as an open position for first-order low-resolution spectroscopy. The imaging channel filter wheel contains three Kodak Wratten 2 neutral density options—ND3, an ND3 + ND1 sandwich, and an ND1 + S II sandwich—as well an open position and a completely opaque position. An ND1 + Na sandwich was used during first light observations but has since been replaced by the opaque slot. A plane can be defined by the optical axis of the incoming beam from the telescope and the center of the EMCCD. The small mirrors and focusing lenses are positioned such that the axes of the two beams (the light from the front of the slit and the light from the grating) are directed to the mirrors that have centers ~12mm above and below this plane, M3 and M7. These final mirrors break RIPS' preceding "in-plane" design by bringing the two beams out-of-plane to locations just above and below the EMCCD center. As such, the two image planes are tilted ~2.6° with respect to the EMCCD plane. Since the images are only ~1-3 mm high (along the slit), and the focal ratio at the detector is ~$f$/10 for the spectrum and ~$f$/20 for the slit image, these tilted image planes do not present a problem of defocusing along the slit.

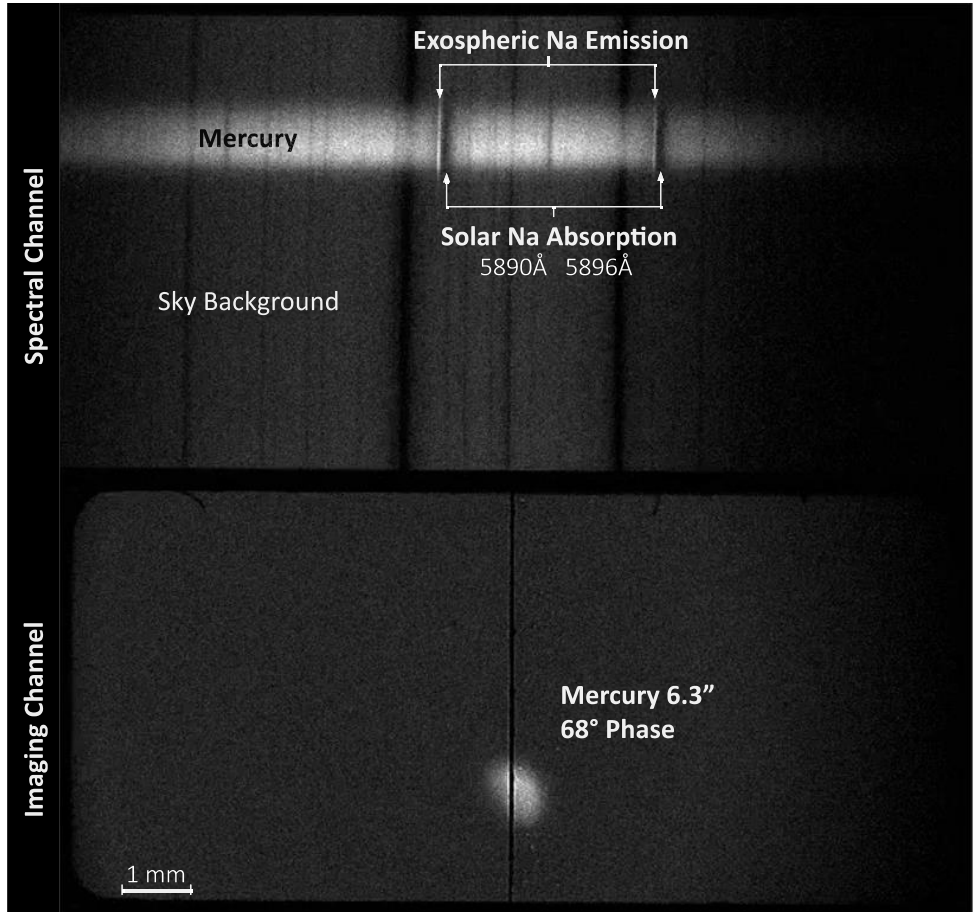

**Figure 2:** An annotated daytime RIPS integration of Mercury with the Na doublet nearly centered. This is a single 7 second integration CCD image with the spectrum recorded on the top half of the detector and the narrowband image on the bottom half. The 6.3" image on the slit in the lower panel is mirrored over the x-axis onto the spectral channel.

Figure 2 shows a raw RIPS frame. The free spectral range here is only 26Å (~0.74° diffracted angle). The EMCCD detector records the spectrum as a dispersed mirror image of the light passing through the slit, magnified by 12%. Each of the 1024 x 1024 square pixels has a 13 μm side length. The maximum pixel readout rate in EMCCD mode is 30 MHz, allowing full-frame image capture at 26fps. Quantum efficiency reaches a maximum of >95% at ~5600Å. Dark current contributes 0.00025 e-/pixel/sec at -80°C, and total system readout noise at 30MHz with no electron multiplying results in 130 e-. With electron multiplying, however, the effective system readout noise is reduced to sub 1 e- levels. The active area pixel well depth is 80,000 e-. Upgrades to the RIPS slit assembly following the first light results in Schmidt et al. (2020) and Lierle et al. (2022) increased the imaging channel area and improved the flatness of the slit jaw's reflectivity.

Instead of a slit decker, RIPS utilizes an assembly with motorized jaws to enable slit width adjustment within the RIPS software. Slit width can range from ~5 μm to 1.8mm, or, equivalently, 0.035" to 12.4" at *f*/55 on a 76 cm telescope. At the narrowest setting, a cold Ne lamp produces linewidths of 2.48 pixels full width at half maximum (FWHM) near the sodium D doublet wavelengths of 5893Å, corresponding to a R~97,000 resolving power. At longer wavelengths, near the 7699Å K D1 lines, a Kr emission source produces a 1.98-pixel FWHM corresponding to a somewhat higher R~126,000. Although it is challenging to fully correct with field flattening, fringing structure inherent to CCD measurements at these wavelengths does not overwhelm the spectral information.

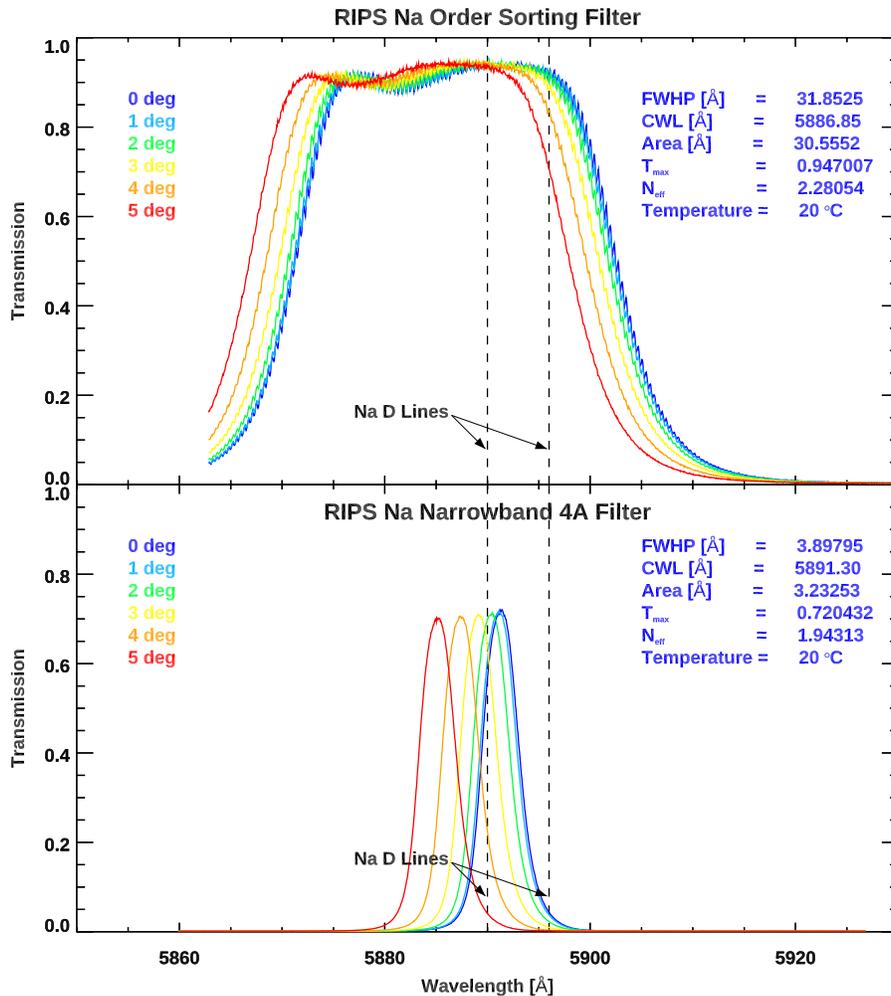

**Figure 3:** The transmission curves of the 32Å RIPS Na order sorting filter and 4Å narrowband Na D2 filter oriented at varying angles to the incident *f*/10 beam. As angle increases relative to the incoming beam, the effective bandpass is shifted blueward.

Unlike most cross-dispersed echelle spectrographs, RIPS was designed for a very long slit length, nearly 6mm, corresponding to 41″ at *f*/55 on a 76 cm telescope. The necessary tradeoff for long slit capability is a narrow free spectral range, and RIPS uses order-sorting filters in its spectral channel filter wheel to isolate a single echelle order. Using the RIPS sodium D-line configuration as an example, Figure 3 shows the transmission curves of an order-sorting filter and an imaging channel filter as a function of incidence angle. Incidence angle considerations are important for any work with narrowband filters. To flux-calibrate the imaging channel, the transmission at line center and area under the filter curve are necessary quantities, and these vary with incidence angle. At the slowest $f$/10 beam speed that RIPS accommodates, the calculated characteristics appear in blue text. At the Perkins Telescope in Anderson Mesa, AZ, which delivers an f/17.5 beam to RIPS, the 4Å narrowband Na filter bandpass is well aligned with the D2 line with no tilt applied. On nearly afocal telescopes, such as the $f$/200 Advanced Electro Optical System in Haleakala, HI, we place an $f$/11 telescope in front of the beam and optimize the throughput by tilting the filter 2-3°. In practice, the wide 40Å Na filter seen in Figure 3 is used as a pre-filter to prevent off-band light from entering the instrument, while the narrow 4Å filter in the imaging channel filter wheel captures a Na D2 image of the target body on the CCD. Each time the beam of light encounters a filter surface inside of RIPS, off-band light may scatter within the optical path. Placing a wavelength-redundant sodium filter at the aperture of RIPS blocks this off-band light from ever entering the instrument. As seen in Figure 3, the transmission of the Na order sorting filter is near 95%, so throughput suffers only slightly when using both filters.

**II.2 Mechanical Design**

Reusing the same collimator and fold lens allows for a compact design. Figure 4 shows the rendered RIPS housing alongside a photograph as-built. Enclosed in an anodized 6061 aluminum housing with a 15.25 cm diameter flange on the front, RIPS measures 71.75 cm along its optical axis, 49.5 cm wide and 20.75 cm tall. A motorized cage encasing RIPS, dubbed the "rotiserizer", rotates the instrument about its optical axis, allowing the slit to align to any desired orientation on sky. The front plate of this assembly is 71 cm x 71 cm with two bolt circles machined to match the Cassegrain focus mounts of the 1.8m Perkins Telescope and the 2.1m Otto Struve Telescope. The central plate of the cage is the widest point at 91.5 cm in diameter with a flat chord to clear the right ascension gear, and the rear cage plate diameter is 50.75 cm. The total end-to-end length along the optical axis of RIPS with the rotiserizer is 99 cm. A high-resolution potentiometer in a voltage divider rotates alongside RIPS to register the instrument's rotation angle and enable repeatable settings within the RIPS control software.

The RIPS slit is aligned with the celestial declination axis during telescope installation. On an equatorially mounted telescope, slewing the telescope in declination identifies this axis. On an altitude-azimuth mount, the slit is aligned perpendicular to the sky drift with telescope tracking disengaged. The capability to quickly change slit position angles proves critical for some planetary science applications. For instance, spectra of cometary comae with alignment along the sunward vector illustrate down tail fall-off rates and orientation of the slit normal to the limb achieves altitude profiles of the lunar exosphere. Aligning the slit along Mercury's spin axis ensures the legitimacy of any measured asymmetries between the north and south hemispheres. Io's tilted plasma torus rocks back and forth with Jupiter's rapid rotation and so position angle rates can be set to allow longer integrations of its relatively faint (< kR) ion emissions.

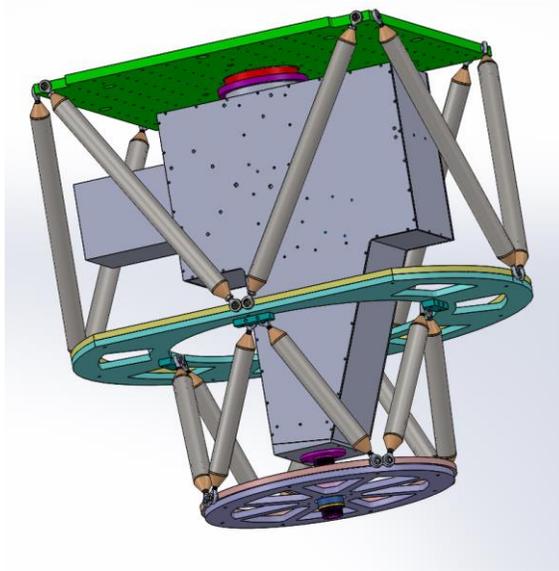 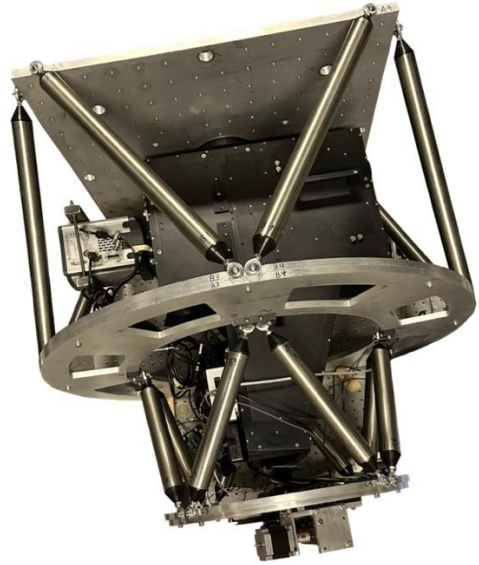

**Figure 4:** *Left:* A SolidWorks rendering of the RIPS housing in its rotating cage known as the 'rotiserizer.' *Right:* RIPS as built.

## II.3 Control Software

The RIPS control software is written in Python 3[3]. The user interface offers full control of all instrument settings aside from those of the EMCCD, which are performed within Andor's proprietary Solis I application. Both the RIPS control software and Solis I can be run on a 32- or 64-bit installation of Windows 7 or newer. Figure 5 displays the main control window. The four onboard motors—slit width, imaging channel focus, spectral channel focus, and diffraction grating angle—are all configured here, as are the two filter wheel positions and the on-sky position angle set by the rotiserizer. Motors interface with the software via COTS Phidget I/O boards. The control software interprets voltage readouts from all motors besides the slit width motor to display the current position of each motor on the interface. The filter wheels communicate and provide registration of their position via serial port. The grating also sends a 0V or 5V binary signal to indicate whether the echelle or first order grating is installed. This information is critical for the 'Spectral Wavelength Adjustment' section of the software interface, which provides a dropdown menu of spectral line settings: Na D, K D1, O I 6300Å, Hα, O I 5577Å, O I 7774Å. Table 1 lists the optimal diffraction orders that correspond to each of these settings. When a feature is selected, the grating rotates to center its wavelength on the detector at the optimal order. Since spectral focus varies with wavelength, the spectral focus motor then tunes the empirically determined focus for the selected feature. This automatic focus routine offers a good starting point, but optimal focus may be affected by ambient conditions such as temperature and pressure. For this reason, it is important to verify focus (either with a cold lamp or test pattern at infinity), and to acquire calibrations in the same conditions as science data. Wavelengths not included in the dropdown menu can be measured by driving the grating motor manually with the wavelength shift buttons. When driving manually, the focus setting is unaffected. The optimal order and corresponding grating angle can be calculated for a user-entered wavelength via Configure > Get Optimal Echelle Angle.

---

[3] RIPS Control Software Version 1.2 available at: https://github.com/plierle/RIPS-Software

| Spectral Line | Order # |
|---|---|
| O I 5577Å | 102$^{nd}$ |
| Na D 5893Å | 97$^{th}$ |
| O I 6300Å | 90$^{th}$ |
| Hα 6563Å | 87$^{th}$ |
| K D1 7699Å | 74$^{th}$ |
| O I 7774Å | 73$^{rd}$ |

**Table 1:** Notable spectral lines studied by RIPS and their corresponding optimal efficiency diffraction orders as determined from the blaze angle of the echelle grating.

In addition to managing the filter wheels and five motors, the RIPS software periodically saves log files with all the instrument's configurations. Once a target data directory has been specified, and a new data file appears there, the instrument configuration logs are automatically read at the file's timestamp and written into the FITS file headers generated by Solis I. These save states can also be created and restored, allowing for precisely repeatable instrument configurations. While the individual filters in the two filter wheels and pre-aperture position do not interface directly with the software, the user is prompted to register them when the software is first run so that this information, too, may be stored in logs, saves, and FITS headers. Logging instrument settings at regular intervals and storing this information in the FITS headers is crucial for RIPS analysis. In particular, when measuring the spectral linewidth on sky, the instrumental linewidth must be calibrated in an identical setup, which can be restored from a save state.

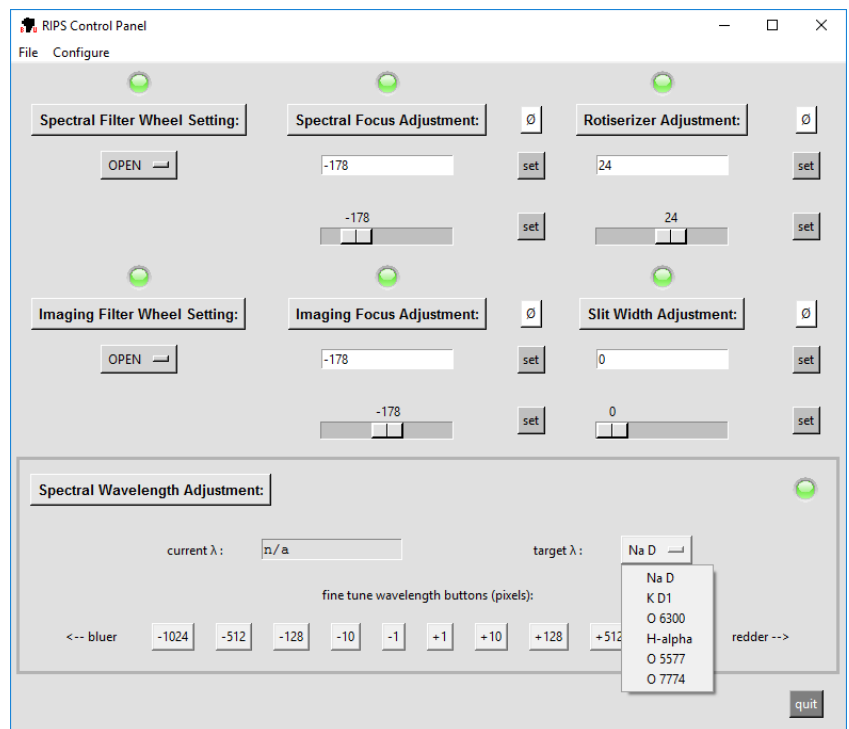

**Figure 5:** The RIPS control software graphical user interface. The two filter wheels can be controlled at the top left, while focus, slit width, and position angle motors can be adjusted using sliders and entry boxes at right. The "Spectral Wavelength Adjustment" section allows the user to center a spectral feature on the EMCCD and adjust in the listed pixel shift increments.

## III. Flux and Wavelength Calibration

Photometric modelling of a body's reflectance spectrum is central to the RIPS data analysis. Mercury and the Moon analysis employs a model produced by Hapke's basic formulation (Hapke 2012; Equation 12.55) and matched to the RIPS instrumental platescale. For Mercury, this formulation uses parameters interpolated from the MESSENGER Mercury Dual Imaging System (MDIS) data set by Domingue et al. (2016). For the Moon, Hapke parameters derive from the Lunar Reconnaissance Orbiter Wide Angle Camera (Sato et al. 2014). For both Mercury and the Moon, the monochromatic slit image captured by RIPS provides frame-by-frame metrics for pointing and seeing, which are applied to rotate and blur the model to best match the observation. By combining this reflectance model with solar spectral irradiance, adjusted for target-Sun geometry and atmospheric conditions, we produce a theoretical model of the continuum brightness at each pixel at a given wavelength. This model is the basis for determining the empirical brightness of exospheric emissions at Mercury and the Moon. Figure 6 shows a narrowband image captured by RIPS, along with the corresponding photometric model before and after blurring for atmospheric seeing.

From this model of continuum brightness in absolute units of Rayleighs / Å, we can empirically determine an instrumental sensitivity in DN Rayleighs$^{-1}$ s$^{-1}$ for a given spectral line. To achieve this, a 1 Å region of continuum is processed alongside the integrated brightness of the targeted emission line. This results in a continuum image of Mercury at the same wavelength as the photometric model. The sensitivity in DN Rayleighs$^{-1}$ s$^{-1}$ is the brightness of the recorded continuum divided by the photometric model. Dividing the exosphere maps reconstructed from the spectral channel by this sensitivity flux calibrates them. Since both the continuum and photometric model are spatially integrated over the entire disk, this calibration is independent of the effective seeing, at least at the global level. At the Dunn Solar Telescope, instrumental sensitivity at 5893Å is determined to be around $6.67 \times 10^{-6}$ DN R$^{-1}$ s$^{-1}$.

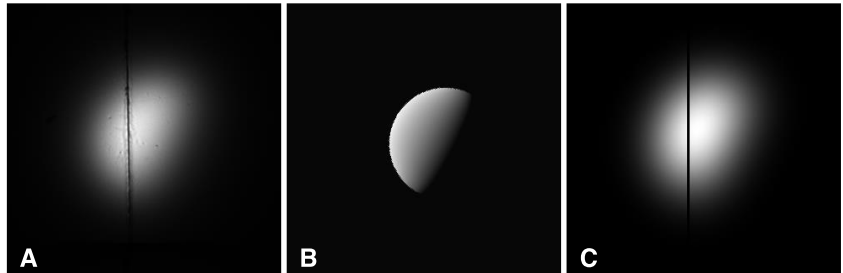

**Figure 6:** (a) A raw narrowband image of Mercury reflected by RIPS's mirrored slit jaws. (b) A photometric model of Mercury's surface reflectance matched to the observing geometry. (c) The photometric model, blurred to match atmospheric seeing, superimposed with a re-creation of RIPS's slit. (Adapted from Lierle et al., 2022)

Flux calibration differs for the Galilean satellites and the Io plasma torus, where Jupiter's disk serves as the reference. Solar irradiance and Jupiter's albedo are both well known (Karkoschka, 1998; Kurucz, 2005) and the reflectivity of the planet's equatorial zone is relatively stable, varying ~3% longitudinally (Chanover et al., 1996), with possible temporal evolution up to 5–10% (Mendikoa et al., 2017). Instrumental sensitivity can be determined using the surface brightness of Jupiter as a photometric reference and applying it to the signal measured from the target satellite, resulting in an absolute brightness for measured emissions there.

Alternatively, wide-slit measurements of photometric standard stars can achieve spectral channel flux calibration, but this technique has not yet been applied since Mercury has been the sole science target yet analyzed in detail. Haze layers near the horizon preclude accurate extrapolation of standard star sensitivity in very high airmass, making photometric modelling generally more accurate. Photometric forward modelling, as above, or standard stars may achieve imaging channel flux calibration. A standard star's surface brightness in Rayleigh units is:

$$I = \frac{10^{-6} (4\pi F) A}{\theta^2} \qquad \text{Eqn. 1}$$

where $F$ is its flux in photons cm$^{-2}$ s$^{-1}$ Å$^{-1}$, $\theta$ is the angular width of a pixel in radians and $A$ is obtained by integrating the area under the filter's transmission curve, e.g., 3.23Å for the sodium filter in Fig. 3. For a RIPS measurement in the field, the imaging channel sensitivity is then the star's measured photometric count rate in DN s$^{-1}$ divided by $I$. Lastly, the filter transmission at the line center must be accounted for. Observers employing narrowband filters sometimes neglect this step, but it is critical: a wide, low transmission filter and a narrow high transmission filter would give the same sensitivity, but measure different count rates from line emission. An imaging channel's science frame, divided by sensitivity in DN s$^{-1}$ R$^{-1}$ and the filter's transmission recovers the surface brightness in Rayleigh units.

The spectral channel of RIPS must undergo rectification to allow a wavelength solution at any given location to be applied across the spatial dimension of the slit. This unwarping process corrects for the curvature of spectral lines inherent in the grating equation, as well as the tilt of the image plane relative to the detector. A 2$^{nd}$ order polynomial is fit to an emission line from a calibration lamp (Ne, Ar, or Kr) near the target wavelength. This polynomial is converted into an array of shifts which are then applied to each row of the spectral channel, rectifying the 2D spectrum.

Wavelength calibration for solar system targets is straightforward and so the RIPS design opted not to use internal calibration lamps, e.g., a ThAr hollow cathode. Instead, a high-resolution solar spectrum (initially a Kurucz model and now TSIS; Coddington et al. 2021) is binned to match the spectral dispersion and convolved with a Gaussian of width proportional to the slit width, effectively matching the spectral resolution. This is cross correlated with a Solar System object's reflectance spectrum to find the pixel shift with peak correlation. The object's 2-way Doppler shift with a light time correction is computed from SPICE (Acton, 2018) and applied to determine the rest frame wavelength solution.

**IV. Example Applications in Planetary Science**

First light of the instrument targeted a Mercury solar elongation from the Perkins 1.8m telescope in Anderson Mesa, AZ on March 15, 2018. RIPS was then installed as a visiting instrument at the 3.7 m Advanced Electro Optical System (AEOS) telescope at Haleakala, HI. Datasets from these two facilities allowed a comparison between high cadence "lucky" spectroscopy, and adaptive optics image stabilization. Schmidt et al. (2020) demonstrated comparable results in mapping the structure of Mercury's Na exosphere. Lierle et al. (2022) analyzed potassium spectra from these campaigns and determined that fainter K D1 line emission demanded higher signal to noise, which precludes high cadence observing modes. Both campaigns occurred before the slit assembly upgrade discussed in Section II.I. Practically, this required that the planetary disk be kept slightly closer to the center of the slit jaws, as the original assembly had a smaller reflective surface area. Any nonuniformities present in the original slit jaws were corrected by flat-fielding.

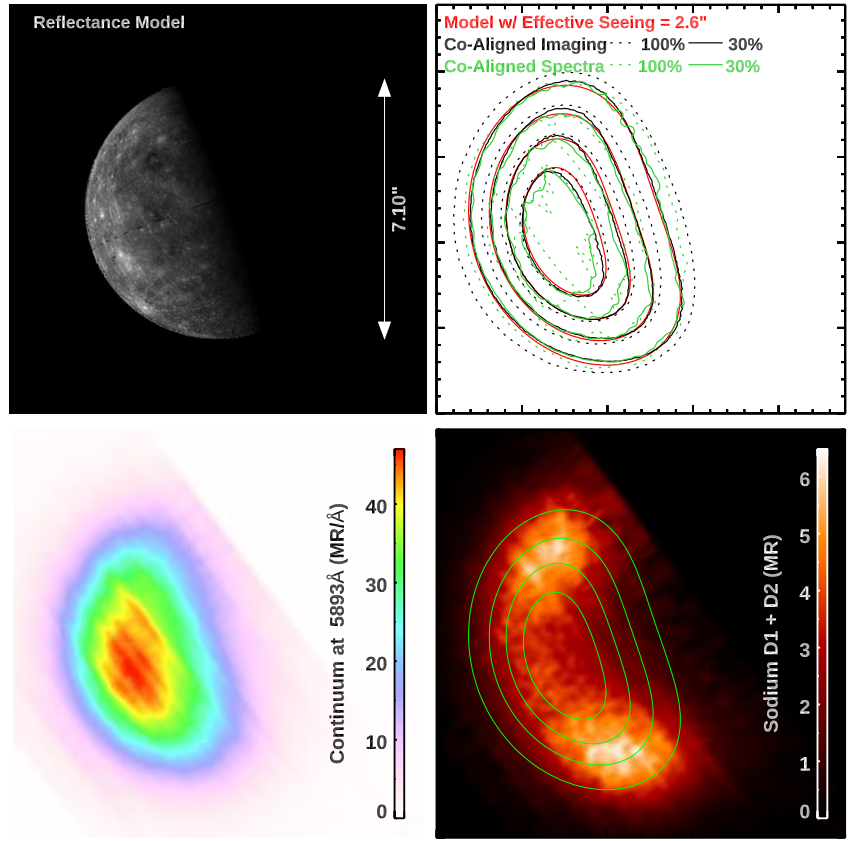

**Figure 7:** Results from RIPS at the AEOS telescope in Haleakala, HI, depicting sodium structure in Mercury's exosphere. Date and time are UTC. The top left panel shows a photometric model with overlaid surface topography. The top right panel shows black contours of the blurred photometric model. Green and red contours indicate the imaging and spectral channels, respectively, with dotted contours constructed from 100% of acquired frames, and solid contours from only the top 30%. The lower left panel shows a reconstructed continuum image of Mercury. The bottom right panel shows Na emission mapped across the disk. These observations were taken before the new RIPS slit assembly was installed. (Adapted from Schmidt et al., 2020)

Figure 7 shows a reconstruction of Mercury's sodium emission from the 2018 AEOS campaign. The top left panel shows the photometric model with overlaid surface topography. The top right panel shows black contours of the photometric model blurred by 2.6"—the mean effective seeing within the sharpest 30% of frames. Green and red contours indicate the imaging and spectral channels, respectively, with dotted contours constructed from 100% of acquired frames, and solid contours from only the top 30%. The lower left panel shows an image of Mercury reconstructed from 1 Å of continuum in the spectral channel. The bottom right panel shows Na emission mapped across the disk. To acquire these spectra, the 40Å Na filter in Fig. 3 was placed in the pre-filter slot, the Na filter was placed in the spectral channel position, and the imaging channel was filtered through a ND1 + Na sandwich. The slit was opened to 15 μm wide and integration times were 1s with electron multiplying enabled.

A similar map was made for potassium using data taken the same night. In this configuration, the pre-filter and spectral channel positions were left open while the imaging channel was filtered through the ND1 + Na sandwich. Differential atmospheric refraction is an important consideration at high airmass, so a ~2" on-sky translation was calculated and applied between potassium spectra and the accompanying narrowband images of Mercury. The slit was kept at 15 μm wide while integration times increased to 5s. K D1 is ~50-100 times fainter than Na D, so spectra had to be combined into larger bins to acquire adequate signal-to-noise. The bin size was chosen for an on-disk SNR ~5. By binning Na to the same resolution and applying the solar photon excitation rate to both species, Lierle et al. (2022) produced the map of the Na/K ratio of column densities shown in Figure 8. This map highlights differences between the Na and K spatial distributions in Mercury's exosphere, particularly at high latitudes. That result was unforeseen since Na and K are generally considered to be chemical analogs (Leblanc et al. 2022).

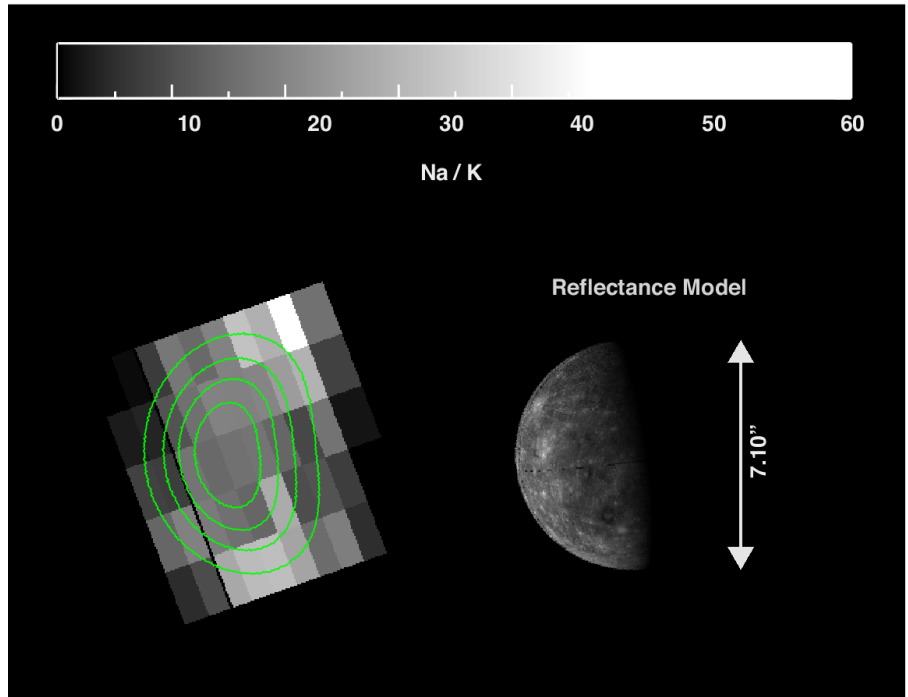

**Figure 8:** RIPS map of the sodium to potassium column density ratio in Mercury's exosphere acquired from the same twilight observation as Fig. 7. 2018 December 12. The reflectance model at right is constructed by overlaying surface topography onto the photometric model. (Adapted from Lierle et al., 2022)

In preparation for BepiColombo, RIPS was deployed for daylight observations at the Dunn Solar Telescope (DST) in Sunspot, NM during a coordinated campaign between several observatories in October 2021. DST is currently the only solar telescope in the United States where pointing restrictions due to off-axis sunlight do not preclude Mercury acquisition. The beam provided by DST was reduced to $f$/55 via a small telescope on the imaging bench before reaching RIPS. A hand paddle allowed fine control of the turret to keep Mercury centered in the imaging channel and to scan the slit across the disk. This proved to be a considerable challenge, however, because shortly after beginning to observe it became clear that the azimuth drive on the telescope was malfunctioning. The result was an unstable image of Mercury bouncing preferentially along the azimuthal axis. The RIPS pipeline cannot easily correct the elongated image of the disk resultant of integration times of a few seconds. Fortunately, the Jovian Interiors from Velocimetry Experiment (JIVE) spectrograph, built to measure Jovian oscillations (Gonçalves et al., 2019), was installed at the DST with a fast tip-tilt system. Using the JIVE front end partially compensated for the telescope's rogue movements and while still not ideal, this fix allowed for some science grade data to be obtained on the last two days of the campaign.

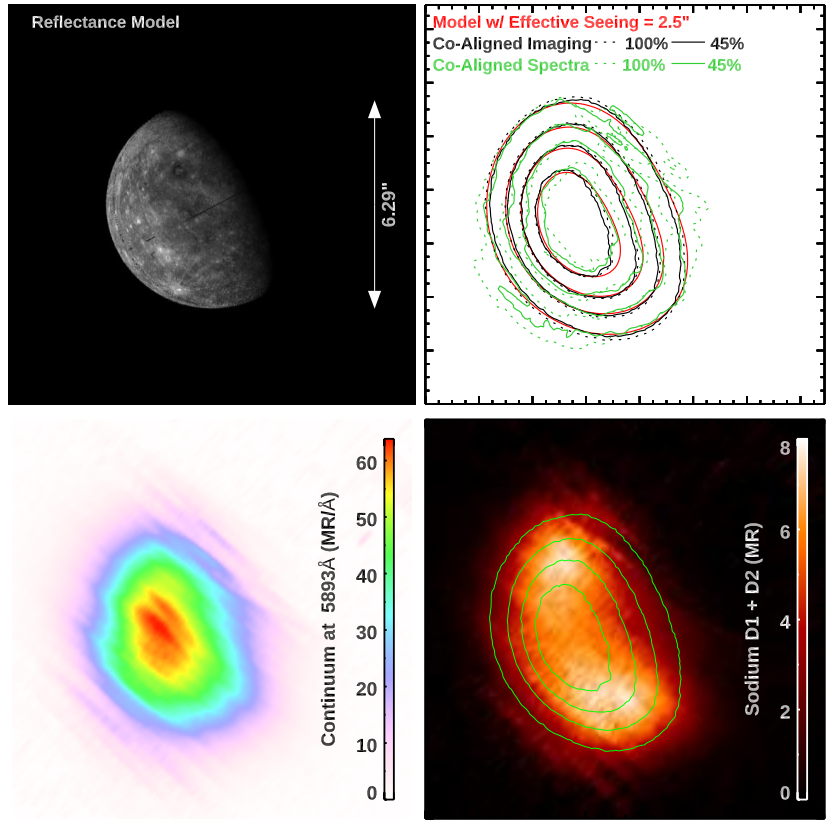

**Figure 9:** Daytime observations with RIPS at DST mapping out exospheric sodium on Mercury. Date and time are UTC. A faulty azimuth drive on the telescope resulted in unstable pointing, resulting in a slightly choppy image of Mercury. These observations were taken with the new RIPS slit assembly installed.

Figure 9 shows results from Mercury observations in daylight on 28 October 2021 at DST under the JIVE fast tip-tilt guider. RIPS was configured with the Na pre-filter and Na spectral channel filters in place and ND3 in the imaging channel. Integration times ranged from 7 to 15s with the slit width set to 10 μm. The effects of unsteady pointing are best seen in the upper right panel, where the improvement from including 100% to including just the luckiest 42% of the frames is material. The measured sodium distribution closely agrees with prior observations showing peaks near the magnetic cusps at mid to high latitudes, and general enhancement in the southern hemisphere when compared to the north. While the reconstructed images of Mercury's continuum and exosphere show rough edges compared to Figure 7, individual frames while the DST azimuth drive was momentarily tracking properly (Fig. 2) show similar quality to those obtained at AEOS and Perkins, giving confidence that future daylight observing campaigns may resolve morphological changes expected in Mercury's exosphere.

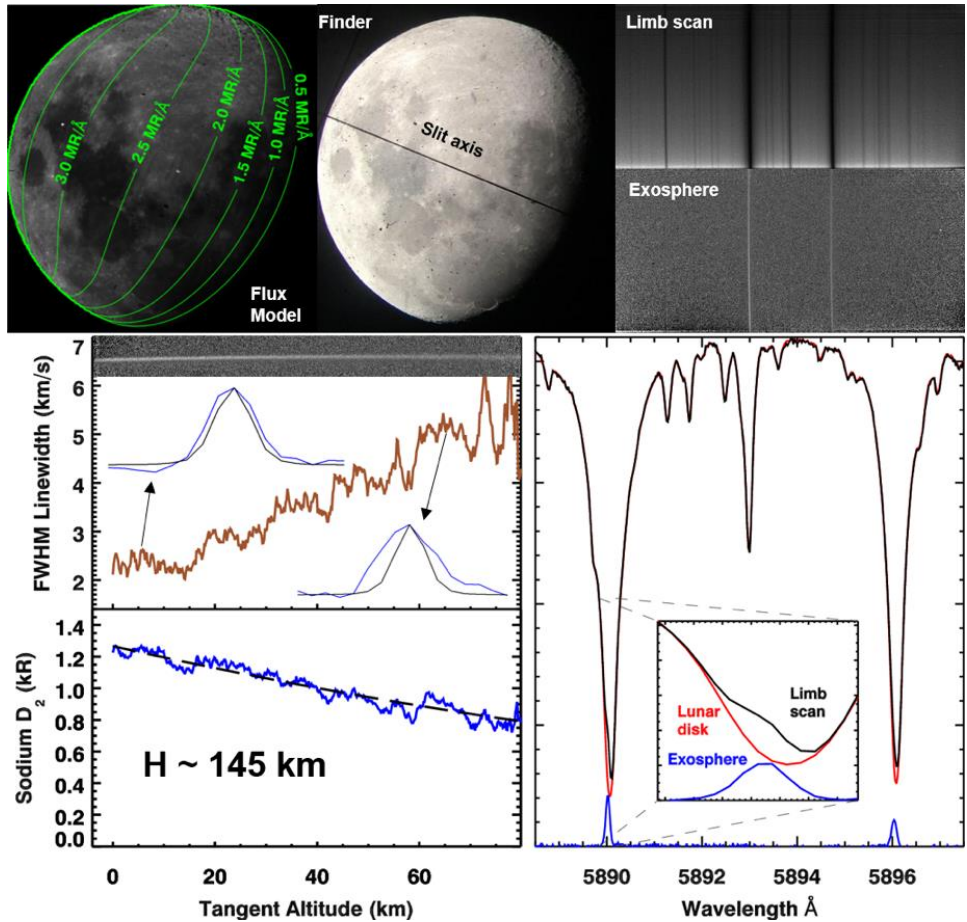

**Figure 10:** Lunar results from April 2018 at the Perkins Telescope showing a broadening Na linewidth with increasing tangent altitude. The top left depicts a photometric model overlaid with surface topography next to an image of the moon from the Perkins finderscope indicating the orientation of the RIPS slit normal to the equatorial limb. On the top right, the spectral channel is shown before (top) and after (bottom) subtraction of the solar reflectance spectrum. The lower panels, from left to right, present the trends observed in Na brightness and linewidth, and illustrate the process of solar subtraction. These data were taken obtained prior to the RIPS slit assembly upgrade.

RIPS targeted the lunar Na exosphere during an April 2018 campaign at the Perkins Telescope in Anderson Mesa, AZ. RIPS was configured with the Na pre-filter and Na spectral channel filters in place and the Na + ND1 sandwich in the imaging channel. The integration time was 600s with the slit width set to 15 μm. The slit was oriented normal to the lunar surface at the equator, as seen in the top middle of Figure 10. Drift scans on the lunar disk provided a spectrum of reflected moonlight that is scaled and subtracted from the limb scan to isolate exospheric emissions, seen in the lower right panel. With the Na D lines isolated, a variation in the linewidth and brightness was evident with tangent altitude above the lunar surface. The bottom left quadrant plots these trends. Brightness fell off with exponential scale height $H$ of ~145 km. Linewidth increased steeply with tangent altitude—a characteristic not previously investigated. Comparison with the line spread profiles of a cold neon emission lamp at slit locations near the surface and at 60 km tangential altitude demonstrates that this is not an instrumental effect. Such broadening of the D lines can be partly attributed to the increase in apex altitudes of more energetic atoms ejected from the lunar surface, and to the escaping population within the Na tail. In the future, we will employ a Monte Carlo model of the linewidth to interpret these characteristics since information on both the gas velocity and brightness versus tangent altitude may constrain the Na ejection rate and velocity distribution function uniquely.

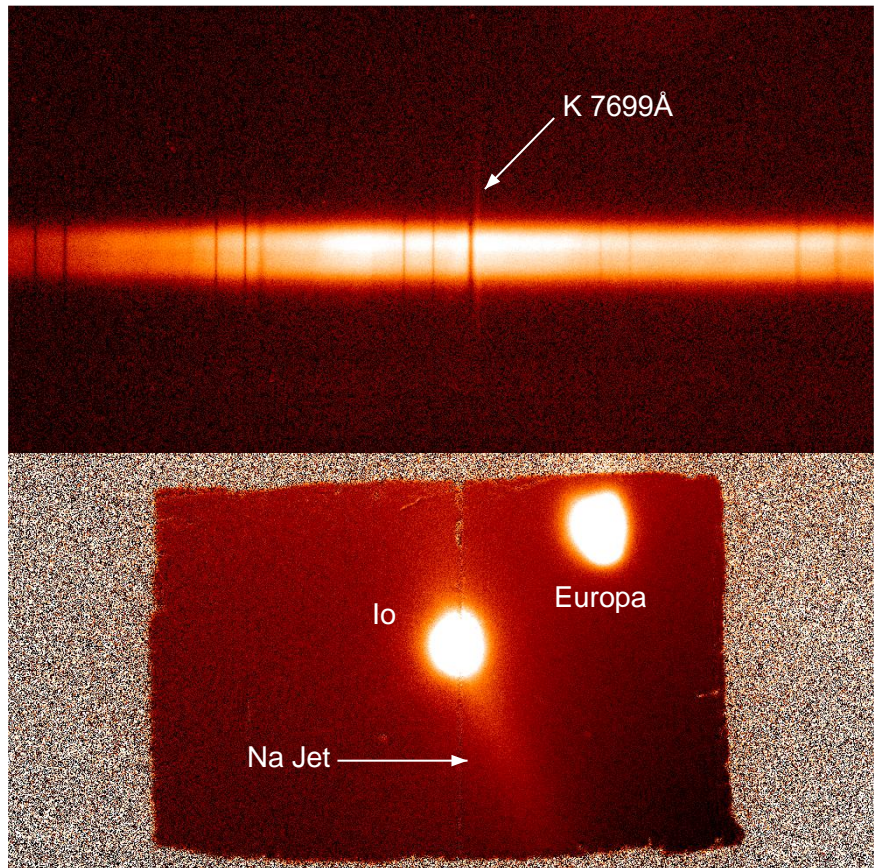

**Figure 11:** A 10-minute integration of Jupiter's moon Io at the Perkins Telescope taken UT 2018 April 27 08:12. In this mode, the Na D2 clouds are imaged while the spectral channel measures K D1. The slit is aligned with Io's orbital plane and ions recombining to form the Na jet illuminate the tilted centrifugal equator of the Io plasma torus. These observations were taken before the new RIPS slit assembly was installed.

Figure 11 illustrates the benefits of narrow-band imaging in concert with high resolution spectroscopy in the case of Io. These data were again obtained with RIPS installed at the Perkins Telescope. While previous examples have made use of filter redundancy to minimize scattered light, this configuration takes advantage of the independent spectral and imaging channel filter wheels to record K spectra while imaging Na. The imaging channel is filtered through the narrowband Na D2 filter in Fig. 3, while the spectral channel and pre-filter slots are left open. This configuration provides spatial context on Io's extended cloud structures, which would not be visible at K wavelengths. This is a single 10-minute integration with the slit width set to 15 μm. The slit was aligned with Io's orbital plane, with up being the direction towards Jupiter, and Europa visible in close proximity. The Na and K cloud in this axis originates in the slow atmospheric loss of neutrals in roughly Keplerian orbits leading and trailing Io. The Na jet is highlighted, tracing ion chemistry from the tilted equator of the Io plasma torus. The rapid disappearance and slow reappearance of Na during Io's passage through Jupiter's shadow was recently interpreted as the interruption of the ion chemical reaction chain:

$$SO_2 + h\nu \rightarrow SO_2^+ + e^-$$
$$SO_2^+ + NaCl \rightarrow NaCl^+ + SO_2$$
$$NaCl^+ + e^- \rightarrow Na^* + Cl^*$$

where the final step yields fast atomic neutrals that spray in a jet from the plasma equator (Schmidt et al. 2023). RIPS data like the 10-minute exposure in Fig. 11 offer valuable insight to the dynamic moon-magnetosphere interaction of Io and its plasma torus. In particular, the jet is predicted to disappear when Io is in Jupiter's shadow due to the interruption of the initial reaction step and slowly re-appear over several hours after egress (Grava et al., 2014a). RIPS could effectively characterize such behavior, which has not yet received a dedicated observation.

The influence of the Io plasma torus extends to its neighboring moon, Europa, which is widely thought to harbor a subsurface ocean. Europa is constantly bombarded by the Io plasma torus, as well as the neutral jets in Io's rapidly escaping atmosphere, resulting in some of Europa's sodium originating at Io. Na sourced from Io would be sputtered only from Europa's trailing hemisphere due to Jupiter's rapid rotation relative to the satellite orbits; however, irradiated NaCl has been detected on the leading hemisphere (Trumbo et al., 2019). Furthermore, the loss rate of Na from Europa is greater than this implantation rate (Leblanc et al., 2005), and the Na/K ratio in Europa's exosphere differs sharply from the clouds escaping Io (Brown, 2001). Together, this offers strong evidence of an endogenic source of alkalis on Europa.

Previous studies suggest that atmospheric Na and K plausibly originate from Europa's subsurface ocean or near-surface brine reservoirs (Leblanc et al., 2002). The surface concentration of NaCl appears highest in Europa's geologically young chaos terrain (Trumbo et al., 2019; 2022), and freezing and over-pressurization of impact melts is thought to yield cryovolcanic eruptions that expel Na-rich water into the atmosphere (Steinbrügge et al., 2020). Salinity is therefore heterogeneous, and since this regulates the freezing point, it is important to identify the amount and location of Europa's subsurface liquids (Chang et al., 2022).

Fig. 12 compares measurements of Na emission lines in Europa's exosphere from RIPS on the 1.8m Perkins Telescope with Keck/HIRES. Na filters were placed in the pre-filter and spectral channel positions, with the Na + ND1 sandwich engaged in the imaging channel. This is a single 1000s integration. The projected slit widths of RIPS and HIRES were 0.30″ and 0.86″, respectively. The 0.30″ slit width reflected naivety in observing Europa, and in the future will be widened to improve SNR. Still, these observations demonstrate the potential for RIPS to acquire science grade data on Europa's Na exosphere.

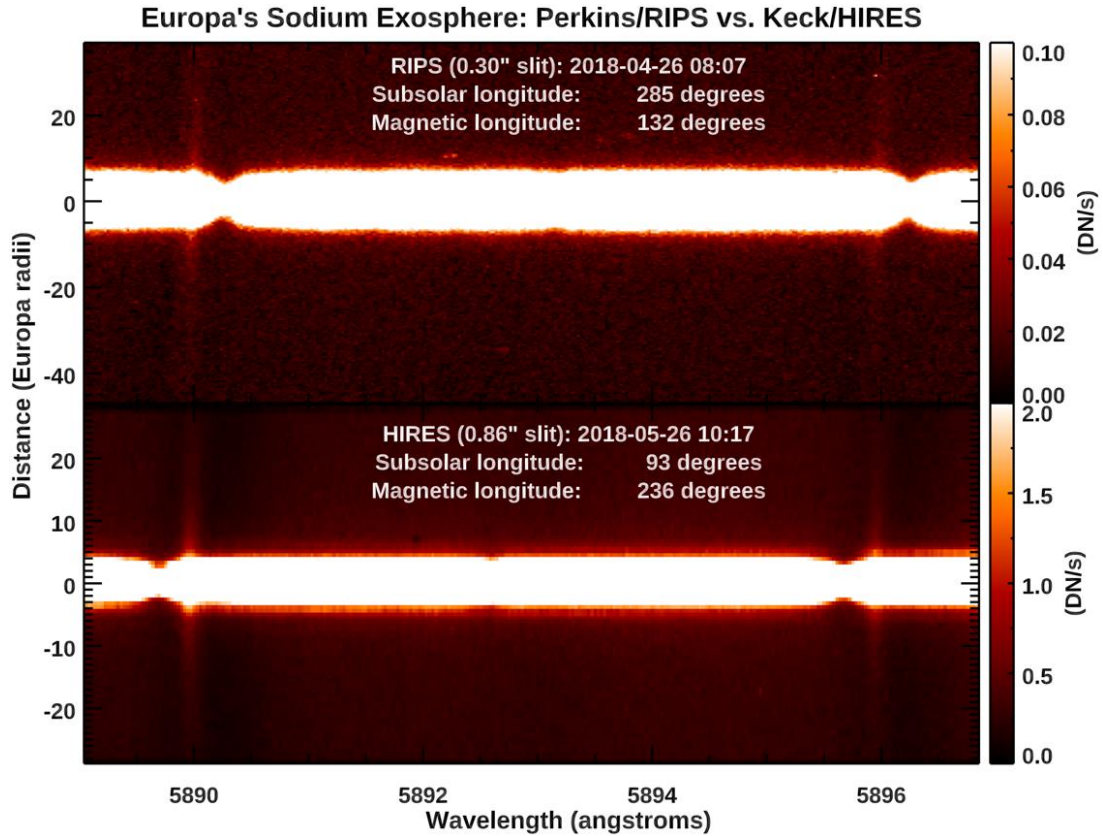

**Figure 12:** Comparison of Europa's sodium D-line emission in April 2018 with RIPS (top panel) and data taken in May 2018 with HIRES (bottom panel). The slit width of RIPS and HIRES were 0.30″ and 0.86″, respectively, and a Na order sorting filter was used for long slit capability in each instrument.

Water plumes that Roth et al. (2014) and others have reported remain controversial, but a growing consensus is that Europa's cryovolcanism is intermittent. A Keck/NIRSPEC survey to detect water vapor on Europa by Paganini et al. (2020) found only 1 instance out of 17 was consistent with a plume, but the geyser's estimated content was substantial: a whopping 2095 ± 658 tonnes of water vapor. Saltwater conducts and salinity can be estimated from the induced magnetic field that Galileo recorded. Hand & Chyba (2007) placed minimum salinity at 12 times less than Earth's oceans, or 0.09% Na by weight. Hence, this plume lofted >4.9 × $10^{28}$ sodium atoms into Europa's atmosphere. Most Na would remain bound as NaCl, but this is >16× Europa's entire exospheric Na content (Leblanc et al. 2005). At least half the time, a given plume's location is exposed to sunlight, where NaCl dissociates into Na in just 26 minutes (Schaefer & Fegley 2005). Photo-dissociation alone then yields a 3.2 ×$10^{25}$ Na atoms/s production rate, which well exceeds the 9.2×$10^{24}$ Na atoms/s background rate that Leblanc et al. (2005) calculated (averaged over Europa's orbit). Therefore, such a plume on Europa would likely produce ample atomic Na measurable by RIPS. Although simplified, this calculation suggests that regular measurements of Europa's sodium exosphere may offer a proxy for identifying and characterizing stochastic cryovolcanism.

**V. Commissioning and Visiting Instrument Plans**

RIPS is being commissioned as one of three facility instruments at the Perkins 1.8m Telescope in Anderson Mesa, AZ. Scheduling of these rotating instruments is coordinated with planetary geometry in mind, particularly bright cometary apparitions, Mercury solar elongations and Jupiter oppositions. For ground-based support of the Europa Clipper mission, Perkins offers an appropriate facility for RIPS to monitor and fully characterize Europa's Na exosphere. Europa's exosphere has been studied exclusively with Keck/HIRES and the image-slicer at the now decommissioned McMath Pierce Telescope, and basic properties like the variation with Jovian local time and magnetic longitude remain unknown. The RIPS mounting plate was also machined to couple with the circular bolt pattern of the McDonald Observatory 2.1m. Visiting instrument campaigns at McDonald particularly benefit Mercury, as two other telescopes on site have strong heritage in Mercury observations, enabling joint observations without the usual frustration of timing and weather between geographically disparate sites. Adaptive optics capability with future RIPS campaigns at AEOS would benefit longer integrations required for Mercury K and Europa Na exospheres by stabilizing these images to improve resolution in spatial mapping.

A major objective for the RIPS project is to provide ground-based support for the BepiColombo mission as a visiting instrument at the Dunn Solar Telescope. The two BepiColombo spacecraft were designed to measure the solar wind and exosphere simultaneously and feature two instruments certain to benefit from ground-based context: MSASI, a Fabry-Perot spectrometer dedicated to Na D2 (Yoshikawa et al., 2010), and PHEBUS, an FUV-EUV spectrometer that features a dedicated channel for the 404 nm potassium doublet (Quémerais et al., 2020). These blue lines were not detected by MESSENGER; however, Vervack et al. (2016) discovered that a Mn triplet at 403 nm does emit within this passband. Therefore, ground-based measurements of the brighter K D lines in the red would be useful to disentangle K from Mn in the upcoming PHEBUS dataset. Support for BepiColombo requires the scheduling flexibility and extended observation windows only possible in daylight, and thus compels a solar telescope. With promising first-light results from the DST, we are confident RIPS can provide this support.

The RIPS User Manual is available at: https://carlschmidt.science/RIPS%20Manual.pdf

**Acknowledgements:**

RIPS was designed, built, and commissioned with the support of the National Science Foundation grant AST-1614903. P.L. acknowledges partial support from the Massachusetts Space Grant Consortium. C.S. acknowledges NASA support from grants 80NSSC21K0051 and 80NSSC19K0790. We thank Dr. Ryan Swindle, Prof. Jeff Kuhn, Lt. Ian McQuaid, and Cody Shaw for their assistance installing RIPS as a visiting instrument at AEOS. Tom Bida and the Lowell Observatory staff provided invaluable help during first light measurements on the Perkins Telescope at Anderson Mesa. We also thank Doug Gilliam and Edward Fish for their support and resourcefulness during our daytime observations at DST, and Prof. Jason Jaciewicz for offering use of the JIVE fast tip-tilt feed.